# On the Mechanism of Light Transmission by Müller Cells

Vladimir Makarov,[1] Lidia Zueva,[2] Igor Khmelinskii,[3*] and Mikhail Inyushin[4]

[1]University of Puerto Rico, Rio Piedras Campus, PO Box 23343, San Juan, PR 00931-3343, USA

[2]Sechenov Institute of Evolutionary Physiology and Biochemistry, Russian Academy of Sciences, St. Petersburg, Russia

[3]Universidade do Algarve, FCT, DQB and CIQA, 8005-139, Faro, Portugal

[4]Universidad Central del Caribe, Bayamón, PR 00960-6032, USA

## *Abstract*

We report that Müller cells contain bundles of long specialized intermediate filaments, each about 10 nm in diameter; most likely, these filaments are the channels providing light transmission to photoreceptor cells in mammalian and avian retinas. We interpret transmission of light in such channels using the notions of quantum confinement, describing energy transport in structures with electro-conductive walls and diameter much smaller than the wavelength of the respective photons. Model calculations produce photon transmission efficiency in such channels exceeding 0.8, in optimized geometry. We infer that protein molecules make up the channels, proposing a qualitative mechanism of light transmission by such structures.

I.    Introduction

* Corresponding author, email: ikhmelin@ualg.pt

Novel nanoscale devices, such as quantum wells (QW) and quantum dots (QD), created new possibilities for development of microelectronics and microoptics. One of the promising approaches is based on photonic crystals allowing the control of dispersion and propagation of light [1,2]. The best-known effects include transmission or rejection of light in a given wavelength range and waveguiding light along linear and bent defects in a periodic photonic crystal structure. A surface plasmon is a transverse magnetic-polarized optical surface wave that propagates along a metal-dielectric interface. Surface plasmons exhibit a range of interesting and useful properties, such as energy asymptotes in the dispersion curves, resonances, field enhancement and localization, high surface and bulk sensitivities, and subwavelength confinement. Because of these attributes, surface plasmons have found applications in a variety of areas such as spectroscopy, nanophotonics, imaging, biosensing and circuitry [3-10]. Plasmon theory was extensively used to interpret phenomena of light focusing at nanoscale [11–34]. Technical applications of optical nano-devices have also been widely discussed [35-74], with several examples of natural nanostructures for light harvesting, transmission and reflection described in plants and bacteria (Vukusic, Sambles, 2003; Damjanovich et al., 2002) [75,76].

Interestingly, light transmission in vertebrate retina is restricted to specialized glial cells, which have many other physiological functions. Franze et al. in their pioneering studies demonstrated that Müller glial cells function as optical fibers (Franze et al., 2007)[77]. They demonstrated that Müller cells transmit visible-range photons from the retinal surface to the photoreceptor cells, located deep under the surface (Franze et al., 2007) [77]. In fact, the retina is inverted and the light projected onto it has to pass through several layers of randomly oriented cells with intrinsic scatterers, before it reaches the light-detecting photoreceptor cells (Hammer et al., 1995, Vos, Bouman 1964) [78,79]. However, the guinea pig retina contains a regular pattern of Müller cells arranged mostly in parallel to each other, spanning the entire thickness of the retina ($\approx$500-800 $\mu$m). Müller cells main cylindrical processes (Schultze, 1866) [80], that span the retina resemble "optical fibers" because of their capacity to transmit light [77]. These cells typically have several complex side branches with functions not related to light transmission, with different morphology (Schultze, 1866; Reichenbach et al., 1989) [80, 81]. Thus, main processes of Müller cells create a way for the light to go through the retina. Here we suggest that main processes of these cells may contain specialized molecular structures functioning like optical waveguides. Therefore, it is important to investigate the morphology of Müller cells main processes and develop a theoretical model of light transmission by such cells.

Numerous materials form dielectric-conductive waveguides, including carbon-based materials with high carrier mobility and conductivity (West et al., 2010) [82]. Presently, we shall interpret light transmission by Müller cells [77] based on the model of light transmission by a waveguide with conductive nanolayer coating. Plasmon theory applies to a dielectric waveguide with metal nanolayer coating [1-10]. This theory uses the quantum electronics approach to analyze the electro-magnetic field (EMF) energy transmission by such a waveguide. A plasmon is understood as a conduction-band electron oscillation along the waveguide surface at the external EMF frequency [3-8,16-21,53-69]. Such oscillations are induced by the EMF at the input end of the waveguide, and then the plasmon energy is transformed back into the EMF at the output end of the waveguide [3-8,16-21,53-69]. Presently, we shall focus our attention on the role of quantum confinement (QC) in light transmission by Müller cells (MC) and nanotubes with conductive coating, where QC operates in the direction normal to the coating. We will analyze the EMF-induced transitions between the discrete states created in the conductive coatings, and the excitation transport from the input end of the waveguide to its output end. We shall demonstrate that excitation transport should operate simultaneously with the state-to-state transitions, as the ground and excited-state wavefunctions span the entire surface of the nanocoating. We shall extend this approach to the analysis of light transmission by the MCs, proposing a mechanism of light transmission by such cells.

II. Experimental Methods and Materials

The presently used histological material had been collected in Russia in 1999 during the study of the bifoveal retina of pied flycatcher (*Ficedula hypoleuca)*. The eyeballs of flycatcher chicks 27 days after hatching were fixed in 3% gluteraldehyde with 2% paraformaldehyde in 0.15 M cacodilate buffer and postfixed with 1% $OsO_4$ in the same buffer. The eyeballs were oriented relatively to the position of the pecten and embedded into Epon-812 epoxy resin. Ultrathin sections 60 nm thick were made on a LKB Bromma Ultratome ultramicrotome (L.K.B. Instruments Ltd, Northampton, UK) and examined on JEM 100B (JEOL Ltd, Japan) electron microscope as it was described before (Khokhlova et al., 2000)[83].

III.  Experimental Results

Müller cells in a Pied flycatcher eye span from the vitreous body, where the light is entering the retina, through the entire retina to the photoreceptor cone cells, and thus the cell structure resembles the one previously described in a guinea pig eye (Franze et al., 2007) [77]. The Müller cell's bodies are located in a specific layer inside the retina, sending principal processes to both of its surfaces. Indeed, the electronic micrograms show that the MC endfeet (inverted cone-shaped zone of MCs adjusted to the vitreous body, see Fig.1, red arrows) form the inner surface lining called the inner limiting membrane (ILM), covering the entire inner surface of the retina. Basal processes of the MCs originating from the endfeet (Fig. 1A, green arrows) protract away from the inner surface, normally to it and essentially parallel to each other (Fig.1A, green arrows) while the processes are spreading around branches of other cells. We suggest that similarly to what had been found in guinea pig (Franze et al., 2007) [77], the light incident from the vitreous body is entering the endfeet (Fig.1A, red arrows) and then the excitation energy flows from the basal process to the cell body and then to the apical processes, where it is transferred to the cone photoreceptor outer segment, as shown on the scheme in Fig.1 C, The diameter of the Müller cell main process in Pied flycatcher may be less than 1 micron (Fig.1 A), therefore these cells working as waveguides should be described as natural nano-optical structures.

<Insert Fig. 1>

We also studied the cytoplasm of Müller cells in the apical and basal processes, and have discovered parallel structures that span the cell cytoplasm all the way through from the inner membrane to the photoreceptors. A higher-magnification insert of a part of basal process of a Müller cell (Fig.1 B) shows this structure within the cytoplasm going within the cell process and repeating all of its curves. This structure resembles the bundle of parallel intermediate filaments with the outer diameter about 10 nm, with some smaller microparticles organized around the filaments. Intermediate filaments in glia cytoplasm are most often associated with globular particles of chaperone proteins (usually termed crystallines) (Perng et al., 2008) [84] that stabilize long filaments or with nucleoprotein particles (Traub et al., 1998) [85]. These filaments span the most part of 400-500 μm of the Müller cell length, from the endfoot to the photoreceptor (outer membrane), thus resembling the previous observations in other species (Reichenbach, Bringmann, 2010) [86]. The filamentary structure was absent in the Müller cell endfeet. Unfortunately, the precise ultrastructure of this intermediate filament construction in MC and the continuity of filaments that may be important for understanding the mechanism of their operation could not be resolved on the equipment used, and will be investigated additionally. Usually intermediate filaments have external diameter varying at 8-13 nm, each filament usually built of 8 protofibrils (Steven, 1990) [87]. Specific intermediate filaments in lens fiber cells are apparently responsible for the optical transparency of these cells (Matsushima et al., 1997; Alizadeh et al., 2003) [88;89]. In our opinion, bundles of intermediate filaments in Müller cells of Pied flycatcher eye should be the structures responsible for transmission of light energy to the photoreceptor cells, and deserve attention of quantum physics.

IV. Theoretical Models

A. Quasi-classical Approach

We already mentioned numerous studies [1-74] that developed the theory of plasmonic quantum electronics and its applications to dielectric-conductive waveguides. Figure 2 presents this model schematically, where only the $E_\parallel$ component of the electric field may produce plasmons, exciting electronic oscillations polarized along the surface of the waveguide.

<Insert Fig. 2>

The efficiency of energy transmission by a waveguide is significantly dependent on the profile of its input zone [3-18,64-74]. Presently, however, we shall not discuss the plasmonic theory in detail, focusing our attention on the more exact quantum approach.

B. Quantum Confinement

Analyzing tubes with nanometer-thick conductive walls, we must take into account the quantum confinement (QC) [82,90,91], appropriate for light transmission by tubes with the diameter smaller than the wavelength of light .

<Insert Fig. 3>

As we already mentioned, the parallel electric field component $E_\parallel$ contributes to plasmonic excitations, while the normal component $E_\perp$ contributes to exciting the discrete states created by QC in the tube nanowalls. Thus, we shall consider the interaction of these field components with the respective quasicontinuum and discrete electronic states, followed by the EMF emission at the other end of the waveguide. The theoretic analysis of the QC in the device shown in Figure 1 is very complex. Therefore, we simplified the system by considering a cylinder with the internal diameter $r_0$, wall thickness $\rho$, and length $l$. The cylinder is linked to a cone, with the wall thickness $\rho Cos(\alpha)$, height $h$, and aperture $2\alpha$. Still, the problem admits numerical solutions only, with the results that we shall discuss next. We shall assume that the wavefunction amplitude is continuous at the surfaces of contact of the cylinder with the two cones (see Fig. 3). Thus, we shall analyze the excitation at the input cone, excitation transport via the cylinder, and the emission at the output cone, for the sake of symmetry.

To evaluate these phenomena qualitatively, we need to determine the quantum states in the cylindrical part using cylindrical coordinates, and the quantum states in the conical part using conical coordinates. It is impossible to solve the Schrödinger equation analytically in a conical system, which we shall therefore analyze approximately. We shall also assume an increased excited state population density in the output conical piece (the right side of the scheme) as compared to the input conical piece (the left side of the scheme). Thus, energy transport in the model system reproduces phenomena occurring in nature, where the excitation ends up on the chromophore molecules contained in the cone photoreceptor cells, with the high excited-state density in the chromophores modelled by that in the output conical piece.

1. Cylindrical Coordinated

We shall solve the eigenstate problem for conductive nanolayers, using the Schrödinger equation in cylindrical coordinates:

$$\Delta = \frac{\partial^2}{\partial r^2} + \frac{1}{r}\frac{\partial}{\partial r} + \frac{1}{r^2}\frac{\partial^2}{\partial \varphi^2} + \frac{\partial^2}{\partial z^2} \tag{1}$$

We shall use boundary conditions equivalent to an axial potential box with infinite potential walls, an acceptable approximation for qualitative analysis. Generally, one presents the Schrödinger equation in cylindrical coordinates as:

$$-\frac{\hbar^2}{2m}\Delta\psi(r,\varphi,z) = E\psi(r,\varphi,z) \tag{2}$$

where

$$\psi(r,\varphi,z) = \psi(r,\varphi)\psi(z) \tag{3}$$

and $m$ is the effective electron mass. Thus, we rewrite (1) as:

$$-\frac{\hbar^2}{2m}\frac{\partial^2\psi(r,\varphi)}{\partial r^2}\psi(z) - \frac{\hbar^2}{2m}\frac{1}{r}\frac{\partial\psi(r,\varphi)}{\partial r}\psi(z) - \frac{\hbar^2}{2m}\frac{1}{r^2}\frac{\partial^2\psi(r,\varphi)}{\partial\varphi^2}\psi(z)$$
$$-\frac{\hbar^2}{2m}\frac{\partial^2\psi(z)}{\partial z^2}\psi(r,\varphi) = (E_{r,\varphi} + E_z)\psi(r,\varphi)\psi(z) \tag{4}$$

Using coordinate separation, we obtain:

$$\frac{\partial^2\psi(r,\varphi)}{\partial r^2} + \frac{1}{r}\frac{\partial\psi(r,\varphi)}{\partial r} + \frac{1}{r^2}\frac{\partial^2\psi(r,\varphi)}{\partial\varphi^2} + \frac{2m}{\hbar^2}E_{r,\varphi}\psi(r,\varphi) = 0$$
$$\frac{\partial^2\psi(z)}{\partial z^2} + \frac{2m}{\hbar^2}E_z\psi(z) = 0 \tag{5}$$

We solve equations (5) in Appendix I, with the result given by:

$$\psi_{k\Lambda k'} =$$
$$= C''_0 \sqrt{\frac{2}{L}} r^\Lambda \left(\frac{1}{r}\frac{d}{dr}\right)^\Lambda \left[\left(\frac{Sin(kr)}{kr} - \frac{1}{kr_0} tg(kr_0) Cos(kr)\right)\right] e^{i\Lambda\varphi} Sin\left(\frac{\pi n}{L} z\right), \quad (6)$$
$$\times e^{\frac{i}{\hbar}(E_r + E_z)t}$$

for the ground state, and

$$\psi_{k_{exc}\Lambda_{exc}k'_{exc}}(t) =$$
$$= C''_0 \sqrt{\frac{2}{L}} r^\Lambda \left(\frac{1}{r}\frac{d}{dr}\right)^{\Lambda_{exc}} \left[\left(\frac{Sin(k_{exc}r)}{k_{exc}r} - \frac{1}{k_{exc}r_0} tg(k_{exc}r_0) Cos(k_{exc}r)\right)\right], \quad (6')$$
$$\times e^{i\Lambda_{exc}\varphi} Sin\left(\frac{\pi n_{exc}}{L} z\right) e^{-\frac{\gamma_r(E_{r,exc})}{2}t - \frac{\gamma_z(E_{z,exc})}{2}t + \frac{i}{\hbar}(E_{r,exc} + E_{z,exc})t}$$

for the excited states, where the quantization caused by the QC in the radial direction is described by

$$k(r_0 + \rho) == Arctg\left[\left(1 + \frac{\rho}{r_0}\right) tg(kr_0)\right] + n\pi \quad (7)$$
$$n = 1, 2, ...$$

Equation (7) may be studied numerically in order to obtain the values of $k$ and hence the energies of the eigenstates. There is also states quantization in the axial direction, although with $L$ much larger than the diameter their energy spectrum is a quasicontinuum:

$$E_z = \frac{\hbar^2 n'^2 \pi^2}{2mL^2} \quad (8)$$
$$n' = 1, 2, \ldots$$

We introduce $\gamma_i(E_i)$, the quantum state widths, determining the excited state relaxation dynamics, including radiative decay. Note that the $E_\parallel$ component causes transitions within the quasicontinuum, while the $E_\perp$ component causes transitions within the radial discrete spectrum. Note also that the eigenvalues for the general wavefunction (6) are:

$$E_{n\Lambda n'} = \langle \psi_{k\Lambda k'}(r,\varphi,z) | \hat{H} | \psi_{k\Lambda k'}(r,\varphi,z) \rangle. \quad (8')$$

Generally, an eigenstate wavefunction involves the entire system; however, we may still evaluate the time evolution of the excited state as described next [91]. To calculate the probability of the excited state prepared in the initial moment of time, we calculate the square of the absolute value of the probability amplitude:

$$P_{exc}(t) = |A_{exc}(t)|^2 \quad (9)$$

The probability amplitude is then:

$$A_{exc}(t) = \langle \psi_{k_{exc}\Lambda_{exc}k'_{exc}}(t=0) | \psi_{k_{exc}\Lambda_{exc}k'_{exc}}(t) \rangle \quad (10)$$

Where,

$$\psi_{k_{exc}\Lambda_{exc}k'_{exc}}(t=0)=$$
$$=C''_0\sqrt{\frac{2}{L}}r^{\Lambda}\left(\frac{1}{r}\frac{d}{dr}\right)^{\Lambda_{exc}}\left[\left(\frac{Sin(k_{exc}r)}{k_{exc}r}-\frac{1}{k_{exc}r_0}tg(k_{exc}r_0)Cos(k_{exc}r)\right)\right] \quad (11)$$
$$\times e^{i\Lambda_{exc}\varphi}Sin\left(\frac{\pi n_{exc}}{L}z\right)$$

and

$$\psi_{k_{exc}\Lambda_{exc}k'_{exc}}(t)=$$
$$=C''_0\sqrt{\frac{2}{L}}r^{\Lambda}\left(\frac{1}{r}\frac{d}{dr}\right)^{\Lambda_{exc}}\left[\left(\frac{Sin(k_{exc}r)}{k_{exc}r}-\frac{1}{k_{exc}r_0}tg(k_{exc}r_0)Cos(k_{exc}r)\right)\right] \quad (12)$$
$$\times e^{i\Lambda_{exc}\varphi}Sin\left(\frac{\pi n_{exc}}{L}z\right)e^{-\frac{\gamma_r(E_{r,exc})}{2}t-\frac{\gamma_z(E_{z,exc})}{2}t+\frac{i}{\hbar}(E_{r,exc}+E_{z,exc})t}$$

Therefore,

$$P_{exc}(t)=e^{-\gamma_r(E_{r,exc})t-\gamma_z(E_{z,exc})t} \quad (13)$$

The phenomenological parameters $\gamma_r(E_{r,exc})$ and $\gamma_z(E_{z,exc})$ are respectively dependent on $E_{r,exc}$ and $E_{z,exc}$, while $\gamma_r(E_r)=0$ and $\gamma_z(E_z)=0$ for the ground state. We shall discuss transverse (in the radial direction) and longitudinal (parallel to the cylinder axis) excited states. Provided the longitudinal excited state is completely relaxed, the probability to find the system in the transverse excited state is:

$$P_{exc}(t) \propto e^{-\gamma_r(E_{r,exc})t} \tag{14}$$

Conversely, if the transverse excited state is completely relaxed, the probability to find the system in the longitudinal excited state is:

$$P_{exc}(t) = e^{-\gamma_z(E_{z,exc})t} \tag{15}$$

We conclude that the conductive nanolayer contains the excitation energy immediately after the excitation, whereupon the excitation decays according to (13) – (15). To analyze the energy transport, we shall next consider absorption and emission at the input and output cones, respectively.

2. Qualitative analysis of processes at the input and output cones

Making use of the axial symmetry of the system, we shall use cylindrical coordinates. Appendix II presents the secular equation (SE) and the respective boundary conditions. It is only possible to solve the SE numerically, although we may obtain some qualitative results even without solving it. Next, we present numerical results for light transmission by the device of Fig. 1.

In the input cone (Fig. 2), the electric field component interacts with the internal conical surface, exciting both longitudinal and transverse states. The energy absorbed upon excitation is proportional to

$$Z_{abs} \propto (S_{in} - S_{out}) \int_0^\infty \frac{\rho_{exc}(\omega)}{\rho_g} E^2(\omega) \left| \langle \psi_g(r,\varphi,z) | \vec{r}_e | \psi_{exc}(r,\varphi,z) \rangle \right|^2 d\omega$$
$$= \pi \left( (h \cdot tg(\alpha) + r_0 + \rho)^2 - r_0^2 \right) \quad (16)$$
$$\times \int_0^\infty \frac{\rho_{exc}(\omega)}{\rho_g} \left| C_{\varphi,g} C_{\varphi,exc} \right| E^2(\omega) \left| \langle \psi_g(r,z) e^{-i\Lambda_g \varphi} | \vec{r}_e(|r_e|,\varphi,z) | \psi_{exc}(r,z) e^{i\Lambda_{exc}\varphi} \rangle \right|^2 d\omega$$

where $E(\omega)$ is the electric field amplitude spectrum of the EMF, $\rho_g = \rho_{g,long}\rho_{g,transv}$, $\rho_{exc}(\omega) = \rho_{exc,long}(\omega)\rho_{exc,transv}(\omega)$ are the level densities of the ground and excited states. Here, we assume that the level density of the ground state is constant, while that of the excited state is dependent on $\omega$; $\psi_g(r,\varphi,z), \psi_{exc}(r,\varphi,z)$ are the wavefunctions of the ground and excited states and $\vec{r}_e(|r_e|,\varphi,z)$ is the vector of the electron location in the conductive nanolayer. The absorbed energy is then transferred to the cylindrical part, here the probability of the excited-state wavefunction transmission ($\xi$ - the transmission coefficient) and reflection ($\zeta$ - the reflection coefficient) depends on the angle $\alpha$. The transmitted energy has a maximum in function of $\alpha$, and therefore may be optimized. Similarly, we may describe the energy transmission to the output cone. The excited states in the output cone will emit with the same spectrum as that of the excitation, as the relaxation of the excited states is quite slow, compared to the excitation transfer in our model (Figure 3). Generally, the transmission and reflection coefficients are, respectively:

$$\xi(\omega) = \xi_0(\omega) \cdot e^{i\varphi_\xi(\omega)}$$
$$\zeta(\omega) = \zeta_0(\omega) \cdot e^{i\varphi_\zeta(\omega)}$$
(17)

Here, $|\xi(\omega)| + |\zeta(\omega)| = 1$; $\xi_0(\omega)$, $\zeta_0(\omega)$ are real functions of $\omega$, and $\varphi_\xi(\omega)$, $\varphi_\zeta(\omega)$ are the phase angles. Thus, the fraction of the absorbed energy transmitted to the excited states of the output cone is:

$$Z_{exc} \propto \pi\left((h \cdot tg(\alpha) + r_0 + \rho)^2 - r_0^2\right)$$
$$\times \int_0^\infty |\xi(\omega) \cdot \xi'(\omega)| \frac{\rho_{exc}(\omega)}{\rho_g}$$
$$\times \left|C_{\varphi,g} C_{\varphi,exc}\right| E^2(\omega) \left|\left\langle \psi_g(r,z) e^{-i\Lambda_g \varphi} \middle| \vec{r}_e(|r_e|,\varphi,z) \middle| \psi_{exc}(r,z) e^{i\Lambda_{exc} \varphi} \right\rangle\right|^2 d\omega$$
(18)

where $\xi(\omega), \xi'(\omega)$ are the transmission coefficients from the input cone to the cylinder, and from the cylinder to the output cone, respectively. If $\xi_0(\omega) = \xi'_0$, $\Delta\varphi_\xi(\omega) = 2\pi\omega L$, and

$$Z_{exc} \propto \pi\left((h \cdot tg(\alpha) + r_0 + \rho)^2 - r_0^2\right)$$
$$\times \int_0^\infty |\xi_0(\omega)|^2 \cdot Cos^2(2\pi\omega L) \frac{\rho_{exc}(\omega)}{\rho_g}$$
$$\times \left|C_{\varphi,g} C_{\varphi,exc}\right| E^2(\omega) \left|\left\langle \psi_g(r,z) e^{-i\Lambda_g \varphi} \middle| \vec{r}_e(|r_e|,\varphi,z) \middle| \psi_{exc}(r,z) e^{i\Lambda_{exc} \varphi} \right\rangle\right|^2 d\omega$$
(19)

Let us consider EMF emission by the output cone. Assuming two equivalent cones, the Pointing vector $\boldsymbol{P'}$ of the output EMF will be parallel to the input $\boldsymbol{P}$, and parallel to the axis of symmetry. Then the density of the emission intensity is:

$$G_{em} = \frac{Z_{exc}}{\pi\left((h\cdot tg(\alpha)+r_0+\rho)^2 - r_0^2\right)}\varphi_{em} \qquad (20)$$

where

$$\begin{aligned}\varphi_{em} &= (\gamma_d \tau_{em})^{-1} \\ \gamma_d &= \frac{1}{\tau_{em}} + \gamma_{relax}\end{aligned} \qquad (21)$$

and $\gamma_{relax}$, $\tau_{em}$ are the radiationless relaxation width and the characteristic emission time, respectively. We assume that $\gamma_{relax} = \frac{2\pi}{\hbar}|\langle V_{ED}\rangle|^2 \rho_D << (\tau_{em})^{-1}$, because $|\langle V_{ED}\rangle|\rho_D << 1$. Here, $\langle V_{ED}\rangle$ is the matrix element of the interaction coupling emitting and dark states in the system, and $\rho_D$ is the density of the dark states. Therefore, $\varphi_{em} \approx 1$. Thus, the energy transmission coefficient is:

$$T = \frac{G_{em}}{G_{abs}} \approx \frac{Z_{exc}}{Z_{abs}} \qquad (22)$$

The latter relationship determines the light transmission by the QC mechanism.

In the following section, we will carry out numerical analysis of the QC effects for a device shown in Figure 4.

<Insert Fig. 4>

3. Numerical Analysis of light transmission by the device of Fig. 4.

Fig. 4 shows the relevant geometrical parameters. The internal diameter of the cylindrical tube is $r_0$, its wall thickness is $\rho$, same as that of the conical pieces. The length of the cylindrical tube is $L_2$, the length the entire device is $L_1$, the length of the input and output pieces is $L_3 = \dfrac{L_1 - L_2}{2}$, $R$ is the curvature radius of the curved conical pieces. We shall assume $r_0 >> \rho$. The larger radius of the input and output cones is:

$$R_d = r_0 + R - \sqrt{R^2 - L_3^2} = r_0 + R\left(1 - \sqrt{1 - \dfrac{(L_1 - L_2)^2}{4R^2}}\right) \quad (23)$$

The boundary ring shown in the zoom-in box in Fig. 4 is normal to both internal and external limiting surfaces.

We shall obtain the numerical solution of the SE in cylindrical coordinates, due to the axial symmetry of the system. To describe the interaction with the EMF, we present the linear momentum as follows:

$$\hat{p} = \hat{p}_e - \frac{e}{c}\hat{A} = -i\hbar\nabla - \frac{e}{c}\hat{A} \tag{24}$$

Here, *A* is the vector potential. The potential must be zero inside the device, thus we rewrite the SE as:

$$\begin{aligned}\hat{H}\psi &= \left(-\frac{\hbar^2}{2m}\Delta - i\hbar\frac{e}{mc}\nabla\cdot\hat{A} + \frac{e^2}{c^2}\hat{A}^2\right)\psi \\ &\approx \left(-\frac{\hbar^2}{2m}\Delta - i\hbar\frac{e}{mc}\nabla\cdot\hat{A}\right)\psi = E\psi\end{aligned} \tag{25}$$

Here, the first term in the Hamiltonian determines the motion of the electron in the nanolayer and the second – its interaction with the EMF. The second term describes a perturbation to the steady-state problem, describing mixing of the ground state with the excited states. Using the perturbation theory, we first obtain the states in absence of a perturbation:

$$\frac{\partial^2\psi(r,z)}{\partial r^2} + \frac{1}{r}\frac{\partial\psi(r,z)}{\partial r} + \frac{\partial^2\psi(r,z)}{\partial z^2} + \left(k^2 - \frac{\Lambda^2}{r^2}\right)\psi(r,z) = 0 \tag{26}$$

Here, $\Lambda$ is the orbital momentum projection on the symmetry axis. The geometry of the device (Fig. 4) in conjunction with the condition of infinite potential on the walls determine the boundary conditions. Using these, we obtain the energies $\{E_{k,\Lambda}\}$ and the respective eigenvectors $\{\psi_{k,\Lambda}(r,z) \cdot e^{i\Lambda\varphi}\}$ describing the unperturbed states. Next, we calculate the probability of the transitions induced by the EMF:

$$W = \frac{2\pi}{\hbar} \left|\langle \hat{V}_{\Lambda_g k_g, \Lambda_e k_e} \rangle\right|^2 \delta(k_e - k_g - k_\lambda)$$
$$= \hbar \frac{2\pi e^2}{m^2 c^2} \left|\langle \psi_{k_g,\Lambda_g}(r,z) \cdot e^{i\Lambda_g \varphi} | \vec{\nabla}_e \cdot \hat{A} | \psi_{k_e,\Lambda_e}(r,z) \cdot e^{i\Lambda_e \varphi} \rangle\right|^2 \delta(k_e - k_g - k_\lambda) \quad (27)$$

Here, $k_g$, $k_e$ and $k_\lambda$ are the wave vectors of the ground and excited states, and of the incident EMF, respectively. We may then use the latter relationship and calculate the transition probability leading to light emission in the output cone of the device.

We present the vector potential as:

$$\hat{A} = \sum_{k_\lambda} \vec{A}_{k_\lambda} e^{i(\vec{k}_\lambda \cdot \vec{r} - \omega_\lambda t)} \quad (28)$$

Here, $k_\lambda = \frac{\omega_\lambda}{c}$. Since the Pointing vector $\vec{P} = [\vec{E} \times \vec{H}]$ points along the device axis, the electric and magnetic field vectors $\vec{E} = -\frac{1}{c}\frac{\partial \vec{A}}{\partial t}$, $\vec{H} = [\nabla \times \vec{A}]$ are normal to it. Thus,

$$(\vec{\nabla}_e \cdot \hat{A}) = \sum_{k_\lambda} \left(\vec{\nabla}_e \cdot \vec{A}_{k_\lambda} e^{i(\vec{k}_\lambda \cdot \vec{r} - \omega_\lambda t)}\right) \quad (29)$$

and

$$W = \frac{2\pi}{\hbar} S_d \left| \left\langle \hat{V}_{\Lambda_g k_g, \Lambda_{exc} k_{exc}} \right\rangle \right|^2 \delta(k_e - k_g - k_\lambda)$$

$$= \hbar \frac{2\pi e^2}{m^2 c^2} S_d \times$$

$$\times \int_0^\infty \left| \left\langle \psi_{k_g, \Lambda_g}(r,z) \cdot e^{i\Lambda_g \varphi} \left| \vec{\nabla}_e \cdot \vec{A}_{k_\lambda} e^{i(\vec{k}_\lambda \cdot \vec{r} - \omega_\lambda t)} \right| \psi_{k_e, \Lambda_e}(r,z) \cdot e^{i\Lambda_e \varphi} \right\rangle \right|^2 \delta(k_e - k_g - k_\lambda) dk_g$$

(30)

where

$$S_d = \pi R_d^2 \tag{31}$$

Note that we substituted the sum over the $k_e$ values in (30) by integration. We used (30) in numerical calculations of the energy transmission efficiency by the device, by the mechanism that involves absorption of a photon at the input cone, and its emission at the output cone. We shall address the spectral selectivity of the device in a follow-up publication.

*Numerical Calculations*

We obtained numerical solutions of the equations (26) and (30) for $\lambda_{EMF}$ = 400 nm, and different parameter values: $\rho$ = 10 nm; $r_0$ = 5$\rho$, 10$\rho$, 15$\rho$, and 20$\rho$; $L_1 = n_1\rho$; $L_2 = n_2\rho$; $L_3 = \rho \frac{n_1 - n_2}{2}$; $R = n_3\rho$ (see Fig. 4). We used the following values of size multipliers: $n_1$ =1000, $n_2$ = 800 and $n_3$ = 100, 200,…, 1000. We present the results for the absorption efficiency:

$$\eta(n_3) = \frac{W}{W_0} \qquad (32)$$

where $W_0$ is the incident EMF intensity within the device cross-section:

$$W_0 = S_d \int_0^\infty \left|E_{k_\lambda}(\omega_\lambda)\right|^2 d\omega_\lambda \qquad (33)$$

We obtained the numerical solution of (26) using the finite differences method [92]. We used homemade FORTRAN code for the numerical analysis of this equation combined with the relationships (30) and (32), with Fig. 5 showing the numerical results.

<Insert Fig. 5>

The efficiency of the light transmission varies with the geometry, with the maximum values corresponding to the following parameter sets: (1) $R = 2.23$ μm, with the maximum cone radius $R_{d,max} = 0.29$ μm; (2) $R = 2.94$ μm, $R_{d,max} = 0.28$ μm ; (3) $R = 3.64$ μm, $R_{d,max} = 0.29$ μm; (4) $R = 4.73$ μm, $R_{d,max} = 0.31$ μm.

4. *Mechanism of Light Transmission by MC*

We found that MC cells contain internal longitudinal channels, with the diameter around 10 nm, of unknown nature. As a hypothesis enabling the usage of the above light transmission mechanism, we shall assume that such channels have electrically conductive walls. Such walls could be built, similarly to carbon nanotubes (CNTs), using conjugated double bonds between carbon and other atoms (N, O, P and S), with carbon being the predominant constituent. Alternative building blocks for the conductive walls that seem more viable biologically are protein molecules, which have high electronic conductivity even in dry state, with values approaching those typical for molecular conductors. These high values of electrical conductivity apparently result from the existence of conjugated bonds in their structure, enabling high electronic mobility [93]. However, in a simplified qualitative approach, we shall now consider the electronic states in a CNT.

*Electronic State Structure of a CNT*

We shall start with a graphene monolayer, which has a conjugated π-system, with the energy gap between bonding and antibonding/conductive zones asymptotically vanishing with increasing system size [94]. Local symmetry at each carbon atom is $D_{3h}$, while the symmetry of graphene elementary cell is $D_{6h}$. Macroscopic symmetry of the graphene sheet depends on its geometry. Presently, we shall analyze the effects of the local $D_{3h}$ symmetry around a C atom. On each C atom, there are three hybridized atomic orbitals

$$\psi_{AO}^{D_{3h}} = \left\{ \begin{array}{c} \frac{1}{\sqrt{2}}\left((2s)+(2p_y)\right) \\ \frac{1}{\sqrt{3}}\left((2s)+\frac{\sqrt{3}}{2}(2p_x)-\frac{1}{2}(2p_y)\right) \\ \frac{1}{\sqrt{3}}\left((2s)-\frac{\sqrt{3}}{2}(2p_x)+\frac{1}{2}(2p_y)\right) \end{array} \right\},$$

located within the plane of the sheet, and one atomic orbital $\left(2p_{z,AO}^{D_{3h}}\right)$, normal to the sheet [95]. These orbitals form an ortho-normalized set of states. In a CNT, the local symmetry around a C atom changes to $C_{3v}$, as three of the orbitals now form a pyramidal structure, with the pyramid height dependent on the CNT radius [95]. Thus, the $\left(2p_{z,AO}^{D_{3h}}\right)$ orbital is mixed with other atomic states $\left(\psi_{AO}^{D_{3h}}\right)$, their contribution increasing when the radius is reduced. The respective perturbation creates splitting between bonding and antibonding/conductive zones, with the energy gap between these zones increasing for smaller values of the CNT radius [96]. Similarly, the presence of O, S, N, P, and other heteroatoms affects the energy gap. The interaction with EMF causes transitions to an excited state, with the excited-state wavefunction distributed over the entire CNT, which will facilitate energy transport between the input and the output cones. Now we have no experimental evidence favoring this mechanism, which could explain the earlier recorded light transmission by Müller cells [77]. However, the Müller cell structure we have described earlier [see. Experimental results] is apparently compatible with the proposed mechanism, with CNTs substituted by protein fibers.

In the following section, we present the results of model calculations with $\rho$ = 0.5 nm, providing a better approximation to the geometry of the optical channels in Müller cells (see Sections II and III).

*Numerical calculations for the device with $\rho$ = 0.5 nm*

We carried out modeling analysis for the following parameter values: $\rho$ = 0.5 nm; $r_0$ = $5\rho$, $10\rho$, $15\rho$, and $20\rho$; $L_1 = n_1\rho$; $L_2 = n_2\rho$; $L_3 = \rho\dfrac{n_1 - n_2}{2}$; $R = n_3\rho$, where $n_1$, $n_2$ and $n_3$ are the size multipliers. Note that this time the size multiplies have larger numeric values, so that the total lengths are similar: $n_1 = 20000$, $n_2 = 19960$ and $n_3 = 100, 200, 300, 400, 500, 600, 700, 700, 800, 900,$ and $1000$. Fig. 6 shows the numerical results. Here, all of the parameters $\rho$, $r_0$, $L_1$, $n_1$, $L_2$, $n_2$, $L_3$, $R$, and $n_3$ were defined above earlier.

<Insert Fig. 6>

Once more, the EMF transmission efficiency has a maximum for each of the parameter sets, corresponding to the following values: (1) $R = 95.7$ nm, $R_{d,max} = 3.0$ nm; (2) $R = 125.6$ nm, $R_{d,max} = 5.4$ nm; (3) $R = 155.5$ nm, $R_{d,max} = 7.8$ nm; (4) $R = 213.5$ nm, $R_{d,max} = 10.2$ nm. As it follows from the results, the optimal radius of the input and output sections ($R_{d,max}$) of the optical channels increases in comparison to the radius of the cylindrical section, while the difference $R_{d,max} - r_0$ decreases, with increasing $r_0$. In effect, we obtain $\frac{R_{d,max}}{r_0} - 1 \ll 1$, with the overall geometry very similar to that of a cylindrical tube. However, the most notable result is the high efficiency of the EMF transmission by such small-diameter channels, as obtained in the calculations.

V. Discussion

We analyzed in detail the role of QC in the transmission of electromagnetic energy by axisymmetric nanochannels with conductive walls, with channel diameter much smaller than the wavelength of the electromagnetic radiation. Note that the transmission of the EMF by such structures may be described by the well-developed and frequently applied plasmon - polariton theory [1-37]. The presently developed approach amounts to an extension of the plasmon - polariton theory in the QC limit [31,73]. Indeed, we describe the quantum states in a waveguide built of nanostructured conductive materials. We assume that the interaction of the EMF with the waveguide produces a transition into the quantum excited state that is delocalized over the entire device, in agreement with the standard plasmon-polariton theory approach, where the energy of the EMF is transported within the waveguide in the form of excited-state electrons [1-8,42-72]. However, there are some terminological differences; here the excited state may be described as an exciton, whereas the plasmon - polariton theory describes such excited states as plasmons or polaritons. We presume that these are different designations for the same physical phenomena, resulting from different qualitative descriptions of the behavior of the system.

Presently we analyzed a waveguide device with the same geometry of the input and the output sections. In principle, the energy transfer efficiency may depend on the geometry of each of these sections, thus allowing extra opportunities to optimize it. The optimum size should also depend on the wavelength of the EMF, with a possibility for separate optimization for each of the color-vision photoreceptors. We shall explore these issues in a follow-up publication.

Our main goal was to develop an understanding of the mechanisms of light conductivity by Müller cells; next, we shall discuss the possible chemical composition of the waveguide structures. Note that protein molecules are universal structural units in animal cells. Apparently, appropriate proteins may form the conductive walls of the nanostructured waveguides, as proteins have high electronic conductivity even in dry state, with values approaching those of typical molecular conductors. These high values of protein electrical conductivity probably result from the existence of conjugated bonds in their structure, enabling high electronic mobility [94]. We were unable to find any publications describing the molecular composition of the waveguide structures in Müller cells, therefore these ideas will remain a hypothesis requiring future testing.

## VI. Conclusion

In the present study we report that Müller cells have long channels (waveguide structures), around 10 nm in diameter, spanning the larger part of the cell process, from the apex of the basal endfoot to the photoreceptor cells. Following the idea that such channels act as waveguides, we developed a QC model of light transmission by waveguides much thinner than the wavelength of the light quanta. We used our model for qualitative analysis of light transmission by such waveguides, with the calculations showing that such devices transmit light with high efficiencies from their input to the output section, already in a device with a much-simplified geometry. The mechanism was extended to light transmission by the waveguides (specialized intermediate filaments) in Müller cells, concluding that they may be built of biopolymers with enough electric conductivity for the mechanism to be operational, due to the presence of conjugated multiple bonds in their structure.

## VII. Acknowledgements

The authors are grateful for support to V. M. from PR NASA EPSCoR (NASA Cooperative Agreement NNX13AB22A) and support to M. I. from NIH grant G12

MD007583. M. I. is grateful to Drs. S. Skatchkov, A. Savvinov and A. Zayas-Santiago for stimulating discussions on the role of Müller cells in vision.

**Appendix I.**

In equation (AI.1),

$$\frac{\partial^2 \psi(r,\varphi)}{\partial r^2} + \frac{1}{r}\frac{\partial \psi(r,\varphi)}{\partial r} + \frac{1}{r^2}\frac{\partial^2 \psi(r,\varphi)}{\partial \varphi^2} + \frac{2m}{\hbar^2}E_{r,\varphi}\psi(r,\varphi) = 0 \qquad (AI.1)$$

$$\frac{\partial^2 \psi(z)}{\partial z^2} + \frac{2m}{\hbar^2}E_z \psi(z) = 0 \qquad (AI.2)$$

we can separate radial and angular coordinates, provided the wavefunction may be written as:

$$\psi(r,\varphi) = \psi(r)\psi(\varphi) \qquad (AI.3)$$

Thus,

$$\frac{\partial \psi(\varphi)}{\partial \varphi} = i\Lambda \psi(\varphi)$$
$$\psi(\varphi) = C_\varphi e^{i\Lambda \varphi} \qquad (AI.4)$$

and

$$\frac{\partial^2 \psi(\varphi)}{\partial \varphi^2} = -\Lambda^2 \psi(\varphi) \tag{AI.5}$$

Since

$$\frac{\partial^2 \psi(\varphi)}{\partial \varphi^2} + \frac{2m}{\hbar^2} E_\varphi \psi(\varphi) = 0 \tag{AI.6}$$
$$\psi(\varphi) = C e^{\pm i\Lambda\varphi}$$

we can write

$$\frac{\partial^2 \psi(r)}{\partial r^2} + \frac{1}{r}\frac{\partial \psi(r)}{\partial r} - \frac{\Lambda^2}{r^2}\psi(r) + \frac{2m}{\hbar^2} E_{r,\varphi}\psi(r) = 0 \tag{AI.7}$$

Or it can be rewritten as follows:

$$\begin{aligned}&\frac{d^2\psi(r)}{dr^2} + \frac{1}{r}\frac{d\psi(r)}{dr} + \left(\frac{2m}{\hbar^2}E_{r,\varphi} - \frac{\Lambda^2}{r^2}\right)\psi(r) \\ &= \frac{d^2\psi(r)}{dr^2} + \frac{1}{r}\frac{d\psi(r)}{dr} + \left(k^2 - \frac{\Lambda^2}{r^2}\right)\psi(r) = 0\end{aligned} \tag{AI.8}$$

Boundary conditions for infinite walls:

$$\psi(r) = \begin{cases} 0; & r = r_0 \\ 0; & r = r_0 + \rho \end{cases} \quad \text{(AI.9)}$$

Taking into account the boundary conditions, we conclude that the wavefunction is only defined in the $[r_0, r_0 + \rho]$ interval.

**Solution**

The equation for the radial function may be written as:

$$\frac{d^2\psi(r)}{dr^2} + \frac{1}{r}\frac{d\psi(r)}{dr} + \left(\frac{2m}{\hbar^2}E_{r,\varphi} - \frac{\Lambda^2}{r^2}\right)\psi(r)$$
$$= \frac{d^2\psi(r)}{dr^2} + \frac{1}{r}\frac{d\psi(r)}{dr} + \left(k^2 - \frac{\Lambda^2}{r^2}\right)\psi(r) = 0 \quad \text{(AI.10)}$$
$$k^2 = \frac{2m}{\hbar^2}E_{r,\varphi}$$

For $\Lambda = 0$, we can write:

$$\frac{d^2\psi(r)}{dr^2} + \frac{1}{r}\frac{d\psi(r)}{dr} + k^2\psi(r) = 0 \quad \text{(AI.11)}$$

The solution of this equation may be presented as:

$$\psi_{k0}(r) = C_1 J_{k0}(kr) + C_2 Y_{k0}(kr) \quad \text{(AI.12)}$$

where

$$J_{ki}(y) = \sum_{j=0}^{\infty} \frac{(-1)^j}{\Gamma(i+j+1)j!} \left(\frac{y}{2}\right)^{2j+1} \tag{AI.13}$$

$$Y_{ki}(y) = Cos(\pi \cdot i)[J_{ki}(y) \cdot Cos(\pi \cdot i) - J_{k,-i}(y)] \tag{AI.14}$$

are Bessel's functions. For $i = 0$, the Bessel functions may be written as:

$$J_{k0}(y) = \sum_{j=0}^{\infty} \frac{(-1)^j}{\Gamma(j+1)j!} \left(\frac{y}{2}\right)^{2j+1} \tag{AI.15}$$

$$Y_{k0}(y) = Cos(\pi \cdot 0)[J_{k0}(y) \cdot Cos(\pi \cdot 0) - J_{k,0}(y)] = 0 \tag{AI.16}$$

Thus, the solution may be presented as follows:

$$\psi_{k0}(r) = C_0 J_{k0}(kr) = C_0 \sum_{j=0}^{\infty} \frac{(-1)^j}{\Gamma(j+1)j!} \left(\frac{kr}{2}\right)^{2j+1} \tag{AI.17}$$

The latter relationship may be approximately presented as

$$\psi_{k0}(r) = C_0 \left( C_1 \frac{Sin(kr)}{kr} + C_2 Cos(kr) \right) \tag{AI.18}$$

Taking into account the boundary condition $r = r_0$; $\psi_{k0}(r) = 0$, we will obtain

$$\psi_{k0}(r_0) = C_0\left(C_1 \frac{Sin(kr_0)}{kr_0} + C_2 Cos(kr_0)\right) = 0$$

$$C_2 = -C_1 \frac{Sin(kr_0)}{kr_0 Cos(kr_0)} = -\frac{C_1}{kr_0} tg(kr_0)$$

Thus,

$$\psi_{k0}(r) = C_0 C_1 \left(\frac{Sin(kr)}{kr} - \frac{1}{kr_0} tg(kr_0) Cos(kr)\right)$$
$$= C_0'\left(\frac{Sin(kr)}{kr} - \frac{1}{kr_0} tg(kr_0) Cos(kr)\right) \tag{AI.19}$$

Taking into account the other boundary condition $r = r_0 + \rho$; $\psi_{k0}(r) = 0$, we will obtain

$$\psi_{k0}(r) = C_0'\left(\frac{Sin(k(r_0 + \rho))}{k(r_0 + \rho)} - \frac{1}{kr_0} tg(kr_0) Cos(k(r_0 + \rho))\right) = 0$$
$$\frac{1}{(r_0 + \rho)} tg(k(r_0 + \rho)) = \frac{1}{r_0} tg(kr_0) \tag{AI.20}$$

$$k(r_0 + \rho) == Arctg\left[\left(1 + \frac{\rho}{r_0}\right) tg(kr_0)\right] + n\pi \tag{AI.21}$$
$$n = 0,1,2,..$$

The latter equation can not be solved directly, although it is apparent that the energy of the system states is a discrete function in the radial direction.

To solve the equation

$$\frac{d^2\psi(r)}{dr^2} + \frac{1}{r}\frac{d\psi(r)}{dr} + \left(k^2 - \frac{\Lambda^2}{r^2}\right)\psi(r) = 0 \tag{AI.22}$$

we have to present the radial function as follows:

$$\psi_{k\Lambda}(r) = r^{\Lambda} \chi_{k\Lambda}(r) \tag{AI.23}$$

thus, our equation becomes:

$$\frac{d^2\chi_{k\Lambda}}{dr^2} + \frac{\Lambda^2}{r}\frac{d\chi_{k\Lambda}}{dr} + k^2 \chi_{k\Lambda} = 0 \tag{AI.24}$$

Taking its derivative, we obtain:

$$\frac{d}{dr}\left\{\frac{d^2\chi_{k\Lambda}}{dr^2} + \frac{\Lambda^2}{r}\frac{d\chi_{k\Lambda}}{dr} + k^2 \chi_{k\Lambda}\right\}$$
$$= \frac{d^3\chi_{k\Lambda}}{dr^3} - \frac{\Lambda^2}{r^2}\frac{d\chi_{k\Lambda}}{dr} + \frac{\Lambda^2}{r}\frac{d^2\chi_{k\Lambda}}{dr^2} + k^2 \frac{d\chi_{k\Lambda}}{dr} = 0 \tag{AI.25}$$

Substituting

$$\frac{d\chi_{k\Lambda}}{dr} = r\chi_{k\Lambda+1} \tag{AI.26}$$

we obtain

$$\frac{d^3\chi_{k\Lambda+1}}{dr^3} + \frac{\Lambda^2}{r}\frac{d\chi_{k\Lambda+1}}{dr} + k^2 d\chi_{k\Lambda+1} = 0 \tag{AI.27}$$

which is equivalent to the equation obtained above. Since

$$\chi_{k\Lambda+1} = \frac{1}{r}\frac{d\chi_{k\Lambda}}{dr} \tag{AI.28}$$

we may write

$$\chi_{k\Lambda+1} = \left(\frac{1}{r}\frac{d}{dr}\right)^{\Lambda}\chi_{k0} \tag{AI.29}$$

or taking into account the solution for $\psi_{k0}(r) = \chi_{k0}(r)$, we obtain

$$\chi_{k\Lambda} = \left(\frac{1}{r}\frac{d}{dr}\right)^{\Lambda}\psi_{k0}(r) \tag{AI.30}$$

Taking into account

$$\psi_{k\Lambda}(r) = r^{\Lambda} \chi_{k\Lambda}(r) \tag{AI.31}$$

we can finally write

$$\psi_{k\Lambda} = C_0 r^{\Lambda} \left( \frac{1}{r} \frac{d}{dr} \right)^{\Lambda} \psi_{k0}(r)$$
$$= C_0' r^{\Lambda} \left( \frac{1}{r} \frac{d}{dr} \right)^{\Lambda} \left( \frac{Sin(kr)}{kr} - \frac{1}{kr_0} tg(kr_0) Cos(kr) \right) \tag{AI.32}$$

For the $z$ coordinate, we may write

$$\frac{\partial^2 \psi(z)}{\partial z^2} + \frac{2m}{\hbar^2} E_z \psi(z) = 0$$
$$\psi(z) = C_1 e^{ikz} + C_2 e^{-ikz} \tag{AI.33}$$
$$k = \frac{1}{\hbar} \sqrt{2mE_z}$$

Assuming that for $z = 0$, $\psi(z) = 0$, or

$$\psi(z) = C_0 Sin(kz) \tag{AI.34}$$

We may as well assume that for $z = L$, $\psi(z) = 0$, i.e,

$$kL = n\pi$$
$$k = \frac{n\pi}{L} = \frac{1}{\hbar}\sqrt{2mE_z} \tag{AI.35}$$
$$E_z = \frac{\hbar^2 n^2 \pi^2}{2mL^2}$$

$$\int_0^L \psi(z)dz = C_0^2 \int_0^L Sin^2(kz)dz$$
$$= C_0^2 \left( \frac{L}{2} - \frac{1}{4k} Sin\left(\frac{2\pi n}{L}z\right) \Big|_0^L \right) = C_0^2 \frac{L}{2} = 1 \tag{AI.36}$$
$$C_0 = \pm\sqrt{\frac{2}{L}}$$

Thus, we finally obtain

$$\psi_{k\Lambda k'} =$$
$$= C''_0 \sqrt{\frac{2}{L}} r^\Lambda \left(\frac{1}{r}\frac{d}{dr}\right)^\Lambda \left[ \left(\frac{Sin(kr)}{kr} - \frac{1}{kr_0}tg(kr_0)Cos(kr)\right) \right] e^{i\Lambda\varphi} Sin\left(\frac{\pi n}{L}z\right) \tag{AI.37}$$

Appendix II.

The secular equation may be written as:

$$-\frac{\hbar^2}{2m}\frac{\partial^2\psi(r,\varphi,z)}{\partial r^2}-\frac{\hbar^2}{2m}\frac{1}{r}\frac{\partial\psi(r,\varphi,z)}{\partial r}-\frac{\hbar^2}{2m}\frac{1}{r^2}\frac{\partial^2\psi(r,\varphi,z)}{\partial\varphi^2}$$
$$-\frac{\hbar^2}{2m}\frac{\partial^2\psi(r,\varphi,z)}{\partial z^2}=E_{r,\varphi,z}\psi(r,\varphi,z)$$
(AII.1)

where $r$ and $z$ coordinates may be separated from $\varphi$ using the wavefunction written as:

$$\psi(r,\varphi,z)=\psi(r,z)\psi(\varphi)$$
(AII.2)

Using the same approach as in Appendix I, we may write:

$$\frac{\partial^2\psi(r,z)}{\partial r^2}+\frac{1}{r}\frac{\partial\psi(r,z)}{\partial r}+\frac{\partial^2\psi(r,z)}{\partial z^2}+\left(k^2-\frac{\Lambda^2}{r^2}\right)\psi(r,z)=0$$
$$k^2=\frac{2mE_{r,z}}{\hbar^2}$$
(AII.3)

Boundary conditions for this problem may be written as:

$$\psi(r,z)=\begin{cases} 0; & z_1=\frac{r_0}{tg(\alpha)},\ z_2=h+\frac{r_0}{tg(\alpha)} \\ 0; & r_1(z)=z\cdot Sin(\alpha),\ r_2(z)=z\cdot Sin(\alpha)+\rho\cdot Cos(\alpha) \end{cases}$$
(AII.4)

Since the variables $r$ and $z$ do not separate in Equation (AII.3), the problem may be solved only numerically using the boundary conditions (AII.4).

**Figure captions**

Figure 1. A: Endfeet (red arrows) and basal processes (green arrows) of Müller cells in Pied flycatcher retina. B: High magnification insert from A, showing a part of cytoplasmic structure (green arrows) that has parallel linear elements resembling intermediate filaments. This structure spans the cytoplasm from the narrow part of the basal endfoot to the apical end that wraps around cone photoreceptor, in the direction of light transmission. Scale in A and B: 500 nm.
C: Schematic presentation of Müller cells (green arrows) with their endfeet (red arrows) and the cone photoreceptors (R, G, B). The light propagation direction coincides with the red arrows.

Figure 2. Schematic presentation of the input section of the lightguide with a nano-thick conductive wall. **P** is the Pointing vector, **E** is the electric field vector of the incident EMF, **H** is its magnetic field vector.

Figure 3. The model for qualitative analysis. The waveguide is composed of a tube and two cones, with conductive walls.

Figure 4. The model for numerical analysis, showing a cross-section of the waveguide with conductive walls. $R$ is the curvature radius referred to in Fig. 5 and Fig. 6.

Figure 5. Calculated plots of the transmission efficiency $\eta$ on the curvature radius $R$ (see Fig. 4). The values of the variable model parameters are: $\rho = 10$ nm, (1) $r_0 = 5\rho$, (2) $r_0 = 10\rho$, (3) $r_0 = 15\rho$, and (4) $r_0 = 20\rho$.

Figure 6. Calculated plots of the transmission efficiency $\eta$ on the curvature radius R (see Fig. 4). The values of the variable model parameters are: $\rho = 0.5$ nm, (1) $r_0 = 5\rho$, (2) $r_0 = 10\rho$, (3) $r_0 = 15\rho$, and (4) $r_0 = 20\rho$.

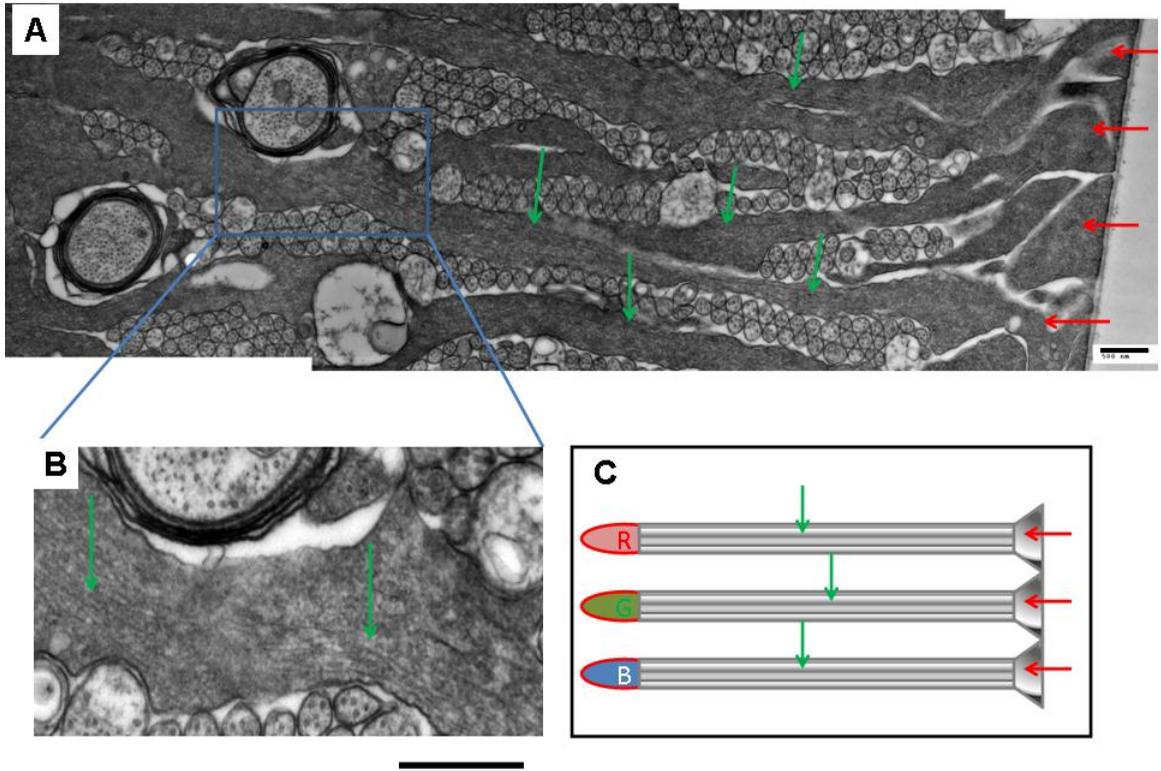

Fig. 1

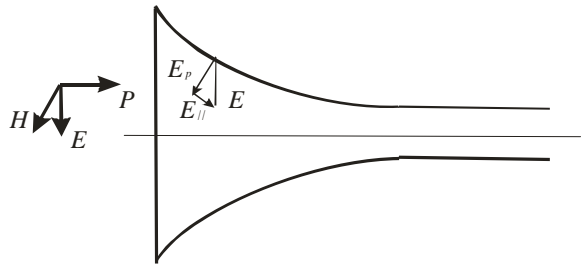

Fig. 2

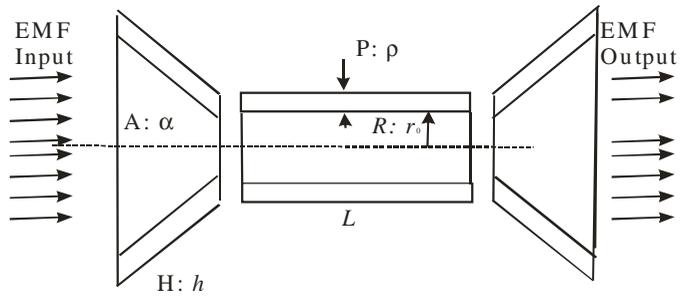

Fig. 3

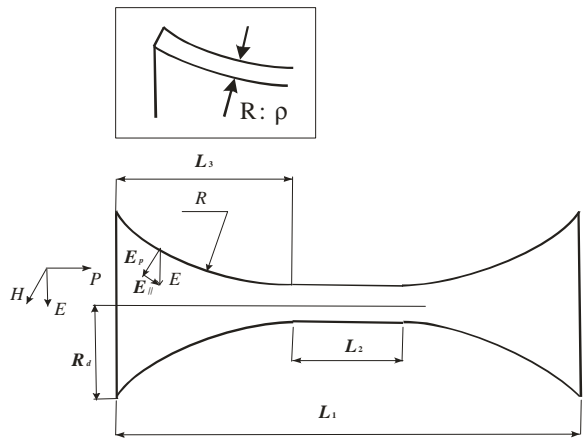

Fig. 4

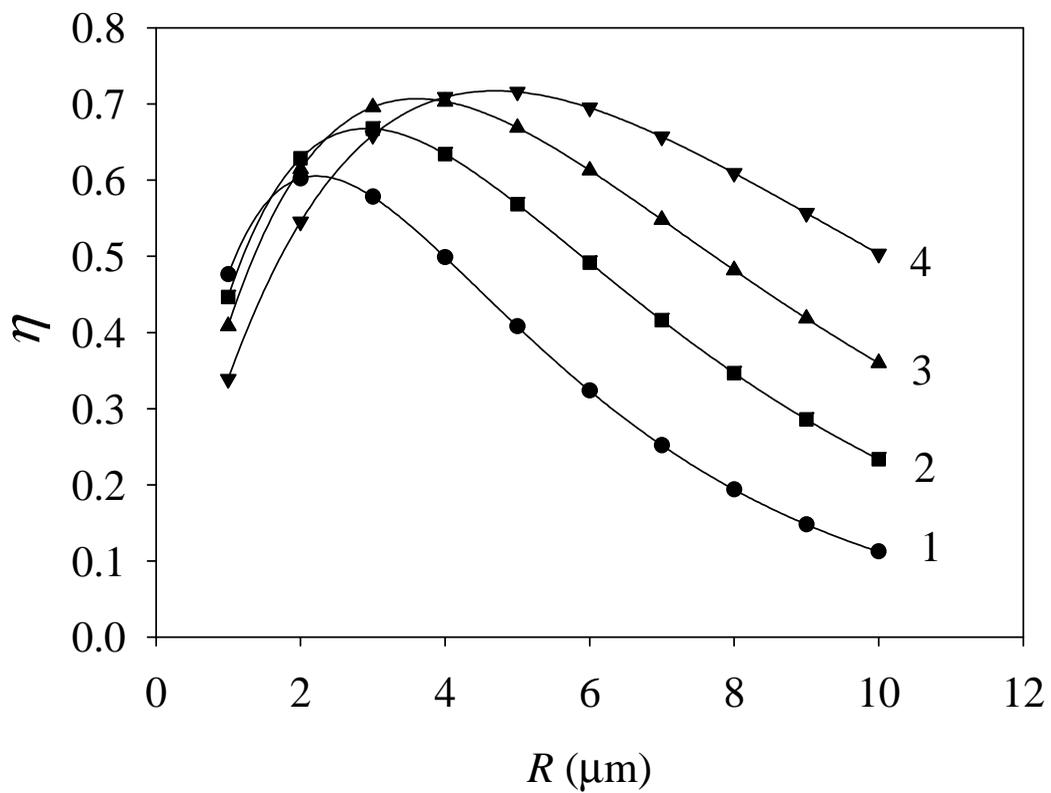

Figure 5. Makarov, et al.

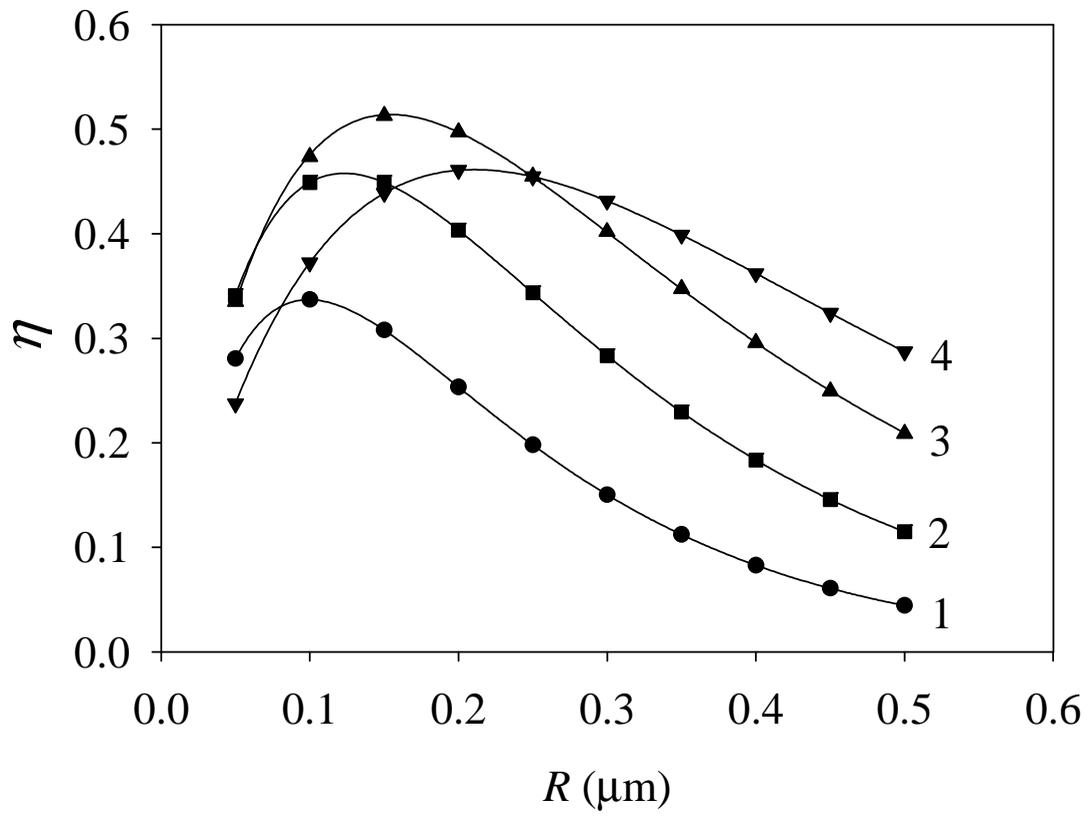

Figure 6. Makarov, etc.